\documentclass[aps,prc,reprint,superscriptaddress,nofootinbib]{revtex4-2}
\usepackage{amsmath,amssymb,booktabs,graphicx,microtype}
\usepackage[hidelinks]{hyperref}
\usepackage{siunitx}
\graphicspath{{figures/}}
\newcommand{\snn}{\sqrt{s_{NN}}}
\newcommand{\pt}{p_{\mathrm T}}
\newcommand{\ncoll}{N_{\mathrm{coll}}}

\begin{document}

\title{Photon-calibrated event-activity bias and subcollision geometry in
\(d+\)Au collisions at \(\snn=200\) GeV with PYTHIA 8 Angantyr}

\author{Muhammad Ajaz}
\email{muhammad.ajaz@cern.ch}
\affiliation{Department of Physics, Faculty of Science, University of Tabuk, Tabuk 47913, Saudi Arabia}

\author{Haifa I. Alrebdi}
\email{hialrebdi@pnu.edu.sa}
\affiliation{Department of Physics, College of Science, Princess Nourah bint Abdulrahman University, P.O. Box 84428, Riyadh 11671, Saudi Arabia}

\author{Muhammad Waqas}
\email{20220073@huat.edu.cn}
\affiliation{School of Mathematics, Physics and Optoelectronic Engineering, Hubei University of Automotive Technology, 442002 Shiyan, People's Republic of China}

\begin{abstract}
Event-activity selections in small nuclear collision systems couple collision
geometry to the soft response accompanying a hard scattering. We examine
this coupling in \(d+\)Au collisions at \(\snn=200\) GeV with PYTHIA~8.316
Angantyr, using direct photons, terminal pre-decay neutral pions, and
anti-\(k_{\mathrm T}\) jets. A hard-bias factor compares the probability for a
hard event to enter an activity class with an
\(N_{\mathrm{coll}}^{\mathrm{ND}}\)-weighted minimum-bias reference. The model
predicts a depleted most-active class and an enhanced peripheral class for
both photons and pions. Photon-tagged events also sample a smaller mean
impact parameter and more nondiffractive subcollisions than pion-tagged or
inclusive HardQCD events. Comparing fully correlated Angantyr events with an
\(N_{\mathrm{coll}}\)-reweighted minimum-bias reference and a factorized
diagnostic separates the explicit geometric contribution from residual
hard--soft correlations. STAR-like and PHENIX-like particle-level activity
proxies preserve the class ordering but give different class probabilities.
The calculation is generator-level and does not identify a unique
microscopic origin for the residual correlations.
\end{abstract}

\maketitle

\section{Introduction}

Classifying a small collision system by forward event activity changes the
sampled impact-parameter and participant distributions. A hard scattering
can also alter the same soft multiplicity used for the classification, so an
activity-selected yield need not follow a Glauber-model
\(\ncoll\) scaling. The geometric basis of this problem is reviewed in
Ref.~\cite{Miller2007}. Its consequences for small-system centrality,
energy-conservation bias, and hard-probe selection have been developed in
Refs.~\cite{PHENIXCentrality2014,Kordell2018,Bzdak2016,Alvioli2013,McGlinchey2016}.
The asymmetric charged-particle distribution measured by PHOBOS further
shows why the rapidity acceptance of the activity estimator matters in
\(d+\)Au collisions \cite{PHOBOS2004}.

RHIC measurements approach the bias from several directions. PHENIX compared
direct photons and high-\(\pt\) neutral pions in activity-selected
\(d+\)Au events \cite{PHENIXPhoton2025} and measured a strong centrality
dependence of reconstructed-jet production \cite{PHENIXJets2016}. STAR
reported the rapidity and species dependence of high-\(\pt\) identified
hadrons \cite{STARRapidity2007}. Identified-particle spectra
\cite{PHENIXIdentified2013} and a system-size survey of neutral-pion
production \cite{PHENIXSystematic2022} provide complementary checks on
fragmentation and event selection.

Direct photons are useful because the prompt component calibrates the hard
scattering without hadronic final-state interactions. The \(pp\) and
\(d+\)Au photon measurements at RHIC establish the corresponding reference
cross sections \cite{PHENIXDirectPP2012,PHENIXDirectDAu2013}; perturbative
calculations describe how prompt-photon production changes in nuclear
collisions \cite{Arleo2011}. A photon--pion comparison does not, by itself,
remove the activity bias. It does make the hard normalization less dependent
on a multiplicity-to-\(\ncoll\) mapping.

PYTHIA provides an exclusive description of the hard process, multiple
parton interactions (MPI), showers, strings, and hadronization
\cite{Pythia82,Pythia83}. Angantyr extends this framework to proton--nucleus
and nucleus--nucleus collisions \cite{Angantyr2018}, building on the
multiple-interaction picture \cite{MPI1987} and Lund string fragmentation
\cite{Lund1983}. Earlier Angantyr calculations of \(d+\)Au collisions
demonstrated the relevance of noncollective contributions to soft and
identified-particle observables \cite{Nayak2026}. The present analysis instead
combines hard probes, two particle-level activity estimators, and stored
event-level subcollision information to distinguish the geometric
contribution from correlations that remain tied to the hard event.

\section{Experimental inputs and observable definitions}

The comparisons use the published PHENIX photon and pion bins
\cite{PHENIXPhoton2025}, PHENIX anti-\(k_{\mathrm T}\), \(R=0.3\) jet bins
\cite{PHENIXJets2016}, STAR high-\(\pt\) rapidity intervals
\cite{STARRapidity2007}, and PHENIX identified-particle spectra
\cite{PHENIXIdentified2013}. Numerical values, bin boundaries, and
uncertainties were read from the associated HEPData records
\cite{HEPData152620,HEPData156988,HEPData101349,HEPData96572}.
Experimental statistical, point-to-point systematic, and normalization
uncertainties remain separate whenever the source provides that distinction.

The direct-photon sample contains photons classified as hard matrix element,
nonhadronic initial-state radiation, or fragmentation final-state radiation.
Photons from hadron decays are excluded, and the residual ``other final''
category is monitored. Neutral pions are counted at their terminal
pre-decay state.

For a minimum-bias hard observable \(X\), we define
\begin{equation}
 Q_{d\mathrm{Au}}^X =
 \frac{\sigma_X^{d\mathrm{Au}}/\sigma_{\mathrm{inel}}^{d\mathrm{Au}}}
      {\langle N_{\mathrm{coll}}^{\mathrm{ND}}\rangle
       \,\sigma_X^{pp}/\sigma_{\mathrm{inel}}^{pp}} .
\label{eq:qdau}
\end{equation}
This dimensionless model quantity uses the realized nondiffractive
subcollision count and is distinct from an activity-independent nuclear
modification factor. Results using the realized total subcollision count are
reported as a normalization variant.

The photon-normalized pion double ratio is
\begin{equation}
 D_{\pi^0/\gamma}^{c} =
 \frac{(\pi^0/\gamma)_{d\mathrm{Au}}^{c}}
      {(\pi^0/\gamma)_{pp}} .
\label{eq:double}
\end{equation}
The imported PHENIX record contains the \(d+\)Au
\(\gamma_{\mathrm{dir}}/\pi^0\) values but not the \(pp\) denominator needed
for Eq.~(\ref{eq:double}). The data comparison is therefore a separately
labelled minimum-bias-normalized diagnostic; no \(pp\) denominator is
inferred.

\section{Generator configuration and event weighting}

Events were generated with PYTHIA~8.316. The \(pp\) tune was selected before
the \(d+\)Au calculation and then transferred without retuning. The Monash
tune \cite{Monash2014} and the Professor methodology
\cite{Professor2010} provide reference points for the parameter and tuning
conventions. HardQCD and prompt-photon processes were generated in exclusive
\(\hat p_{\mathrm T}\) intervals. The nominal settings are summarized in
Table~\ref{tab:generator}.

\begin{table}[t]
\centering
\caption{Generator and analysis definitions used for the reported results.}
\label{tab:generator}
\begin{tabular}{ll}
\toprule
Item & Definition \\
\midrule
Generator & PYTHIA 8.316 Angantyr \\
Systems and energy & \(pp\), \(d+\mathrm{Au}\), \(\sqrt{s_{NN}}=200\) GeV \\
Nominal processes & HardQCD and prompt photon \\
Generated bins & Exclusive \(\hat p_{\mathrm T}\) intervals \\
Direct photons & \shortstack[l]{HARD\_ME + ISR\_NONHADRONIC\\
+ FSR\_FRAGMENTATION} \\
\(\pi^0\) & Terminal pre-decay state \\
Jets & \shortstack[l]{anti-\(k_{\mathrm T}\), \(R=0.3\), E-scheme,\\
visible primaries} \\
Jet constituents & \(p_{\mathrm T}>0.4\) GeV/\(c\), muons included \\
Rapidity convention & Au-going \(y<0\), \(d\)-going \(y>0\) \\
\bottomrule
\end{tabular}
\end{table}

Colour-reconnection alternatives are evaluated separately from the nominal
result. The beyond-leading-colour string model \cite{Christiansen2015} and the
spatially constrained reconnection model
\cite{SCCR2023,SCCR2023Erratum} define the range of model variations
considered. Because the public description of the latter does not uniquely
specify all settings, it is used only to assess the corresponding model
sensitivity and not as an exact reproduction.

Generator comparisons in \(pp\) collisions indicate that charged-particle,
underlying-event, and jet observables constrain different parts of the event
model \cite{Waqar2024}. Accordingly, tune, activity-estimator, and
constituent-definition variations are evaluated separately.

Raw analysis objects were produced with Rivet \cite{Rivet2020}; the
heavy-ion extensions described in Ref.~\cite{RivetHI2020} retain the event
and centrality information needed here. Within each exclusive generated
interval, objects are multiplied by
\begin{equation}
 w_{\mathrm{bin}} =
 \frac{\overline{\sigma_{\mathrm{gen}}}}
      {\sum_{\mathrm{independent\ samples}} W},
\label{eq:weight}
\end{equation}
and compatible generated intervals are then summed. The corresponding
\(\sum W^2\) values are combined with the same weights to propagate MC
statistical uncertainties. The reported results combine 102,000,000 accepted events generated with
independent seeds; the replica contributions are retained separately for
stability and uncertainty checks.

\section{Activity estimators and hard-probe definitions}

The STAR-like estimator counts charged particles on the Au-going side in
\(-3.8<\eta<-2.8\). The PHENIX-like estimator counts particles in a BBC-S
acceptance. Both are generator-level particle proxies; neither includes a
detector response or represents measured BBC charge. Fixed multiplicity
thresholds define classes labelled 0--20\%, 20--40\%, 40--60\%, and
60--100\%.

\begin{table}[t]
\centering
\caption{Weighted hard-event geometry from persisted Angantyr truth.}
\label{tab:geometry}
\begin{tabular}{lrrrrr}
\toprule
Selection & \(\langle b\rangle\) [fm] & \(\langle N_{\rm part}\rangle\) &
\(\langle N_{\rm coll}^{\rm tot}\rangle\) &
\(\langle N_{\rm coll}^{\rm ND}\rangle\) & \(\langle N_{\rm MPI}\rangle\) \\
\midrule
Direct photon & 4.205 & 14.54 & 15.41 & 10.85 & 3.56 \\
Terminal \(\pi^0\) & 4.692 & 11.69 & 11.86 & 7.80 & 3.37 \\
HardQCD process proxy & 4.784 & 11.37 & 11.49 & 7.54 & 3.32 \\
\bottomrule
\end{tabular}
\end{table}

\begin{figure}[t]
\centering\includegraphics[width=\columnwidth]{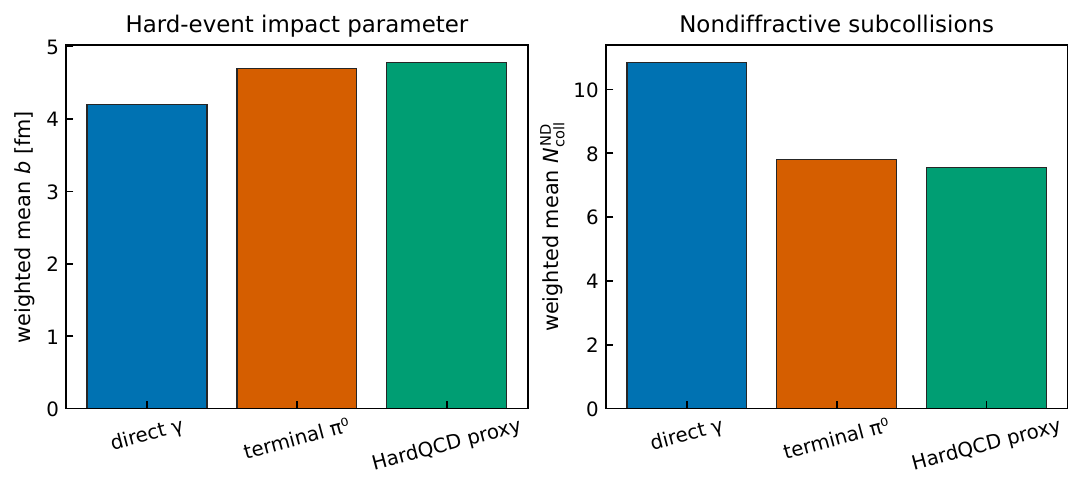}
\caption{Weighted hard-event geometry from Angantyr event truth. Bars show
the weighted means of the impact parameter and the realized
\(N_{\mathrm{coll}}^{\mathrm{ND}}\). The photon selection samples a smaller
impact parameter and more nondiffractive subcollisions than the pion and
inclusive HardQCD selections.}
\label{fig:geometry}
\end{figure}

The stored Angantyr event information gives
\(\langle b\rangle=4.205\) fm and
\(\langle N_{\mathrm{coll}}^{\mathrm{ND}}\rangle=10.85\) for direct-photon
events. Pion-tagged events give 4.692 fm and 7.80, and the inclusive HardQCD
sample gives 4.784 fm and 7.54. Figure~\ref{fig:geometry} displays these
means. These differences characterize the geometry sampled by each hard-event
definition; they do not imply that the probe alters the collision geometry.

Jets are reconstructed with the anti-\(k_{\mathrm T}\) algorithm
\cite{AntiKt2008}, \(R=0.3\), and E-scheme recombination using FastJet
\cite{FastJet2012}. Nominal constituents are visible primary particles with
\(\pt>0.4\) GeV/\(c\); muons are included and invisible particles are
excluded. Variants remove muons, set the constituent threshold to zero, or
retain charged particles and photons only.

The rapidity observable is the charge-summed pion yield at negative rapidity
divided by the yield at positive rapidity. Negative rapidity is Au-going and
positive rapidity is \(d\)-going. The generated intervals from 8 to 9 and 9
to 10 GeV/\(c\) are summed before comparison with the published
8--10 GeV/\(c\) point.

\section{Photon and pion results}

\begin{figure}[t]
\centering\includegraphics[width=\columnwidth]{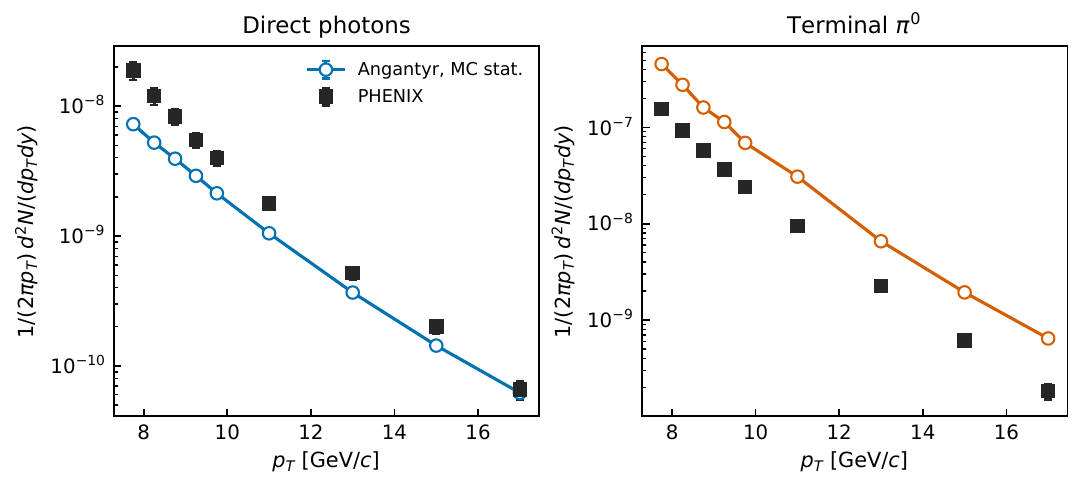}
\caption{Minimum-bias \(d+\)Au direct-photon (left) and terminal pre-decay
\(\pi^0\) (right) invariant yields. Open coloured points show Angantyr with
MC statistical uncertainties. Filled black points show PHENIX values with
the point uncertainties supplied in HEPData.}
\label{fig:spectra}
\end{figure}

Figure~\ref{fig:spectra} shows the minimum-bias spectra. Their MC statistical
precision ranges from 0.58\% to 1.17\% for direct photons and from 2.59\% to
6.36\% for pions. Model and experimental uncertainties are displayed
separately.

\begin{figure}[t]
\centering\includegraphics[width=\columnwidth]{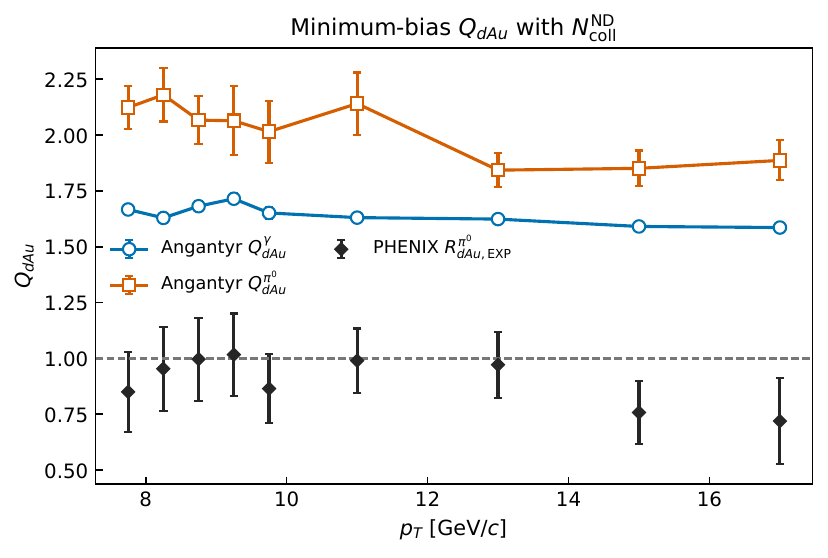}
\caption{Minimum-bias \(Q_{d\mathrm{Au}}\) for direct photons and terminal
\(\pi^0\), normalized by the realized
\(N_{\mathrm{coll}}^{\mathrm{ND}}\). The PHENIX pion quantity uses its
published experimental normalization and is shown for comparison rather
than as an identical model definition.}
\label{fig:qdau}
\end{figure}

With the normalization in Eq.~(\ref{eq:qdau}), the model predicts
\(Q_{d\mathrm{Au}}^\gamma=1.586\)--1.715 and
\(Q_{d\mathrm{Au}}^{\pi^0}=1.843\)--2.180
(Fig.~\ref{fig:qdau}). Values above unity do not, by themselves, identify a
nuclear enhancement: the ratio contains model inelastic cross sections, the
realized \(N_{\mathrm{coll}}^{\mathrm{ND}}\), and the selection implicit in
the hard sample.

\begin{figure}[t]
\centering\includegraphics[width=\columnwidth]{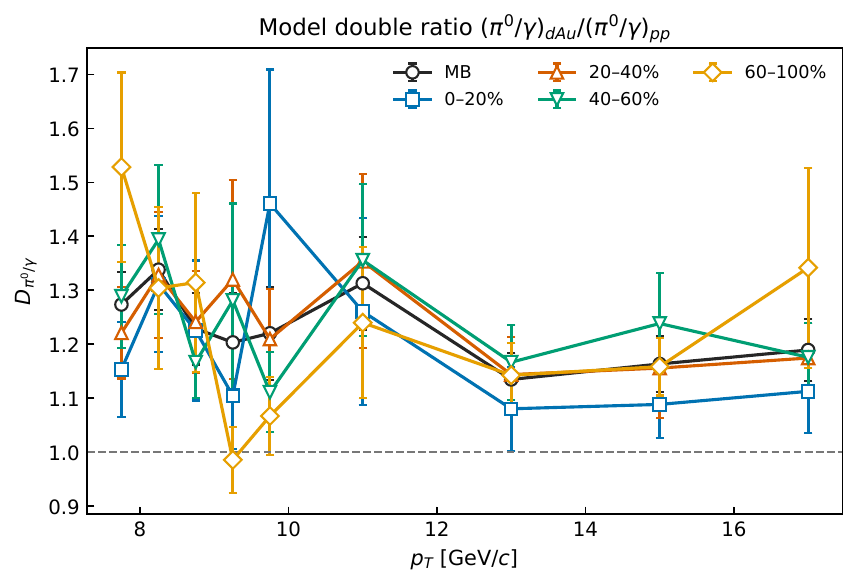}
\caption{Photon-normalized pion double ratio from Eq.~(\ref{eq:double}).
All reported bins are shown. Error bars indicate MC statistical
uncertainties.
The experimental comparison uses a minimum-bias-normalized \(d+\)Au
\(\gamma/\pi^0\) diagnostic, not a constructed \(pp\) denominator.}
\label{fig:double}
\end{figure}

The minimum-bias double ratio lies between 1.135 and 1.338. Across the four
activity classes the range is 0.986--1.528
(Fig.~\ref{fig:double}). Two points in the most-active class are dominated by
MC statistics; they remain visible but do not determine the class trend.

\section{Hard-event bias}

For hard probe \(X\) and activity class \(c\), we define
\begin{equation}
B_X^c =
\frac{P(c\mid X_{\mathrm{hard}})}
{P_{\mathrm{MB}}(c)\,
 \langle N_{\mathrm{coll}}\rangle_c/
 \langle N_{\mathrm{coll}}\rangle_{\mathrm{MB}}}.
\label{eq:bias}
\end{equation}
The nominal definition weights the selected photon or pion candidates and
uses \(N_{\mathrm{coll}}^{\mathrm{ND}}\); an event-triggered alternative is
reported in the Supplemental Material.

\begin{table}[t]
\centering
\caption{Range of the \(N_{\rm coll}^{\rm ND}\)-normalized hard-bias factor
over \(7.5<p_{\mathrm T}<18\) GeV/\(c\).}
\label{tab:bias}
\begin{tabular}{lll}
\toprule
Probe & Activity class & \(B_X^c\) range \\
\midrule
\(B_\gamma\) & 0--20\% & 0.517--0.543 \\
\(B_\gamma\) & 20--40\% & 1.133--1.164 \\
\(B_\gamma\) & 40--60\% & 1.548--1.612 \\
\(B_\gamma\) & 60--100\% & 1.652--1.735 \\
\(B_{\pi^0}\) & 0--20\% & 0.482--0.648 \\
\(B_{\pi^0}\) & 20--40\% & 1.094--1.277 \\
\(B_{\pi^0}\) & 40--60\% & 1.410--1.715 \\
\(B_{\pi^0}\) & 60--100\% & 1.353--1.982 \\
\bottomrule
\end{tabular}
\end{table}

The most active class gives \(B_\gamma=0.517\)--0.543 and
\(B_{\pi^0}=0.482\)--0.648. The peripheral class gives 1.652--1.735 and
1.353--1.982, respectively. For photons, \(B_\gamma\) rises monotonically
from the most-active to the peripheral class. The pion values follow the same overall ordering, with
larger statistical fluctuations. Within the adopted normalization, the class
probabilities therefore depart from an \(\ncoll\)-weighted minimum-bias
expectation. In the model, this pattern arises from event selection relative
to the adopted reference rather than from an imposed final-state suppression
mechanism.

\begin{figure}[t]
\centering\includegraphics[width=\columnwidth]{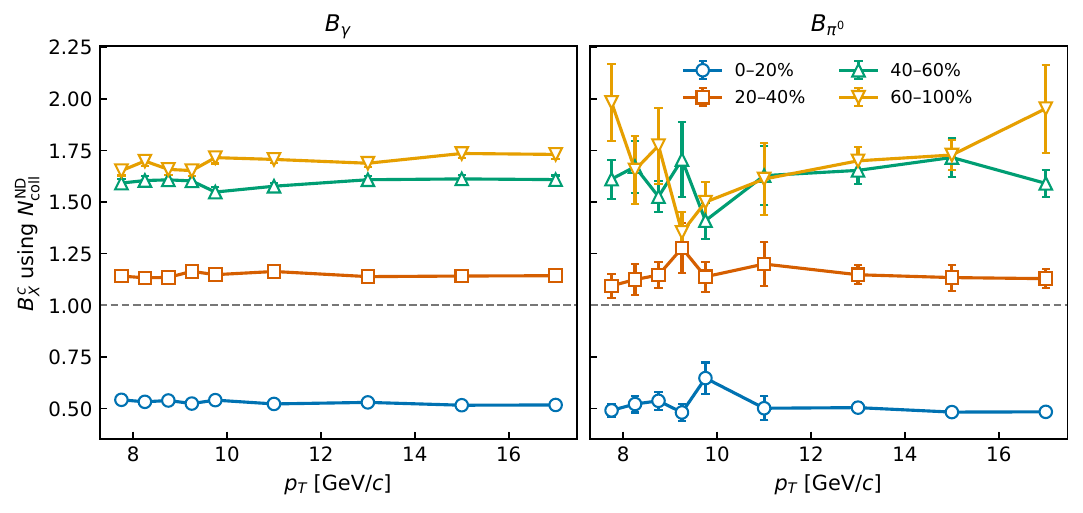}
\caption{Hard-bias factors for direct photons and terminal pions using
\(N_{\mathrm{coll}}^{\mathrm{ND}}\). Unity denotes the
\(\ncoll\)-weighted minimum-bias reference. MC statistical uncertainties are
shown for every point.}
\label{fig:bias}
\end{figure}

\section{Jets and rapidity asymmetry}

\begin{figure}[t]
\centering\includegraphics[width=\columnwidth]{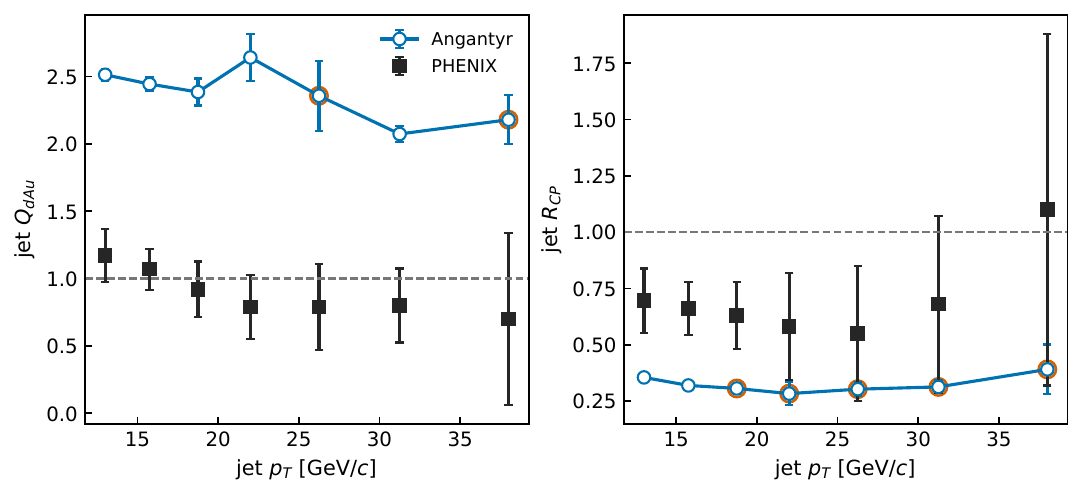}
\caption{Jet \(Q_{d\mathrm{Au}}\) and \(R_{CP}\). Open orange rings mark bins
with larger MC statistical uncertainty. Model and PHENIX uncertainties are
shown separately. The model central and peripheral classes are 0--20\% and
60--100\%; the experimental peripheral class is 60--88\%.}
\label{fig:jets}
\end{figure}

The jet calculation gives \(Q_{d\mathrm{Au}}=1.868\)--2.640 and
\(R_{CP}=0.283\)--0.391. The two \(Q_{d\mathrm{Au}}\) bins and five
\(R_{CP}\) bins with the largest MC uncertainties are shown explicitly in
Fig.~\ref{fig:jets}. Experimental uncertainties exceed the MC uncertainty
in most of these bins. The 34--42 GeV/\(c\) \(R_{CP}\) point has the
largest MC uncertainty; the overall jet conclusion is driven by the
lower-\(\pt\) bins.

\begin{figure}[t]
\centering\includegraphics[width=\columnwidth]{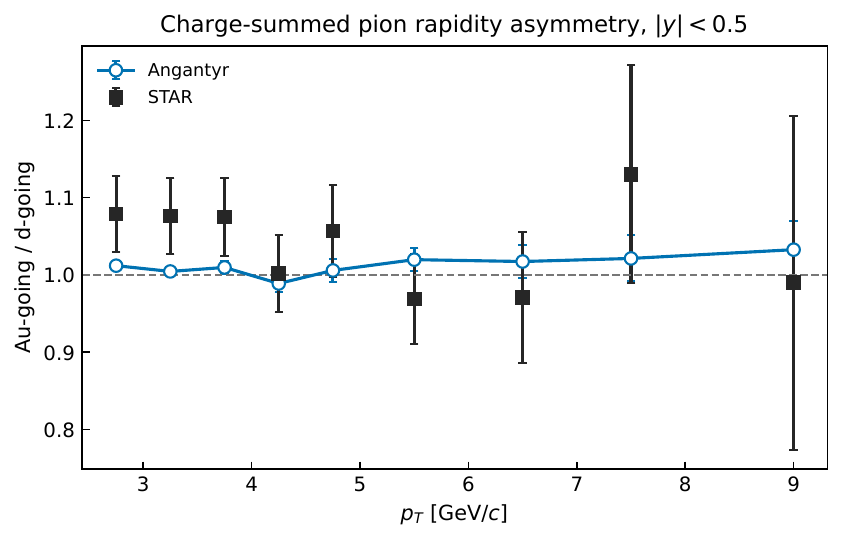}
\caption{Charge-summed pion Au-going/\(d\)-going ratio. Negative rapidity is
Au-going. The 8--10 GeV/\(c\) point is obtained by summing the two adjacent
generated intervals before forming the ratio.}
\label{fig:rapidity}
\end{figure}

The selected charge-summed pion ratio remains between 0.989 and 1.033
(Fig.~\ref{fig:rapidity}), with MC precision of 0.42--3.60\%. It is
consistent with unity and is used as a longitudinal closure check rather
than evidence for a resolved forward/backward asymmetry.

\section{Hard--soft closure and estimator dependence}

Track-based underlying-event comparisons show that the apparent soft
response depends on the hard-event and multiplicity definitions
\cite{Alrebdi2025}. We therefore compare three constructions rather than
assigning the observed class dependence to a single mechanism.

Construction A takes the hard probe and activity from the same Angantyr
event. Construction B uses the minimum-bias class probability reweighted by
the realized \(\ncoll\). Construction C combines the \(pp\) hard process with
a \(d+\)Au soft reference. Construction C is a diagnostic factorized
construction, not a physical event generator.

\begin{figure}[t]
\centering\includegraphics[width=\columnwidth]{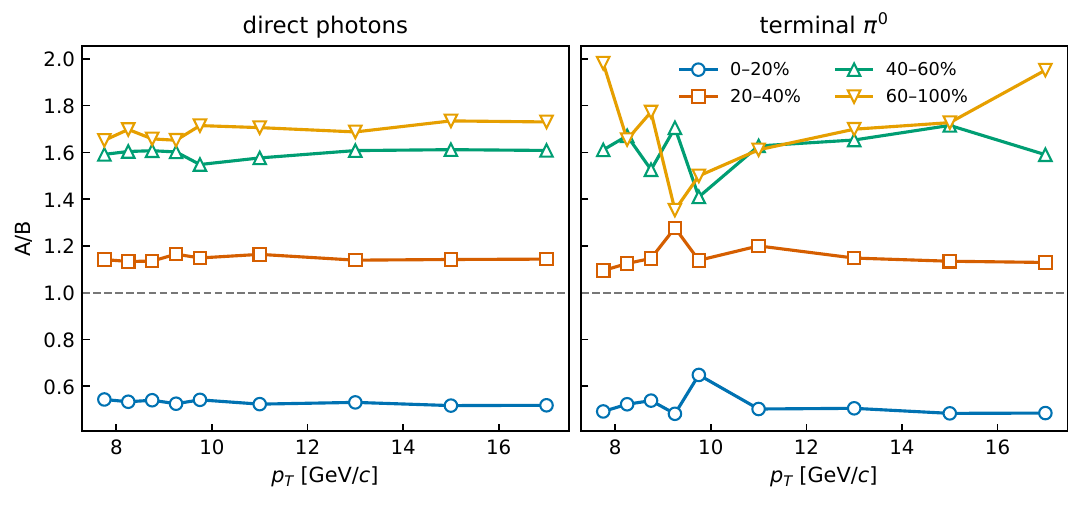}
\caption{Ratio of the fully correlated Angantyr construction (A) to the
\(N_{\mathrm{coll}}\)-reweighted minimum-bias reference (B). The residual
shares the activity-class ordering of the hard-bias factors. This ratio does
not isolate a unique microscopic mechanism.}
\label{fig:closure}
\end{figure}

The three pairwise ratios probe different parts of the response. B/C
isolates the change introduced by the geometry weighting, A/C contains the
full hard-class response relative to minimum bias, and A/B shows the residual
after the explicit geometric reference. Figure~\ref{fig:closure} shows A/B.

The residual may contain contributions from hard-event selection, energy
sharing, MPI, and forward activity. The A/B ratio does not separate these
sources; it quantifies the activity ordering that remains after the explicit
\(\ncoll\)-weighted geometric reference.

\begin{figure}[t]
\centering\includegraphics[width=\columnwidth]{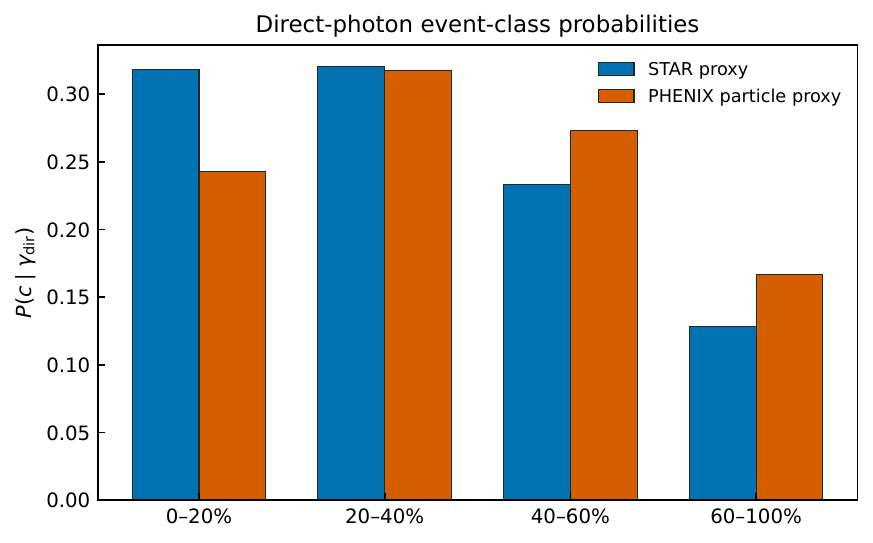}
\caption{Direct-photon class probabilities for the STAR-like and
PHENIX-like particle-level activity estimators. The PHENIX-like quantity is
a BBC-S acceptance proxy, not measured detector charge.}
\label{fig:estimator}
\end{figure}

Changing the activity estimator shifts the numerical class probabilities
without reversing their ordering. For direct-photon events, the most-active
probability is 31.8\% for the STAR proxy and 24.3\% for the PHENIX particle
proxy. The corresponding peripheral probabilities are 12.9\% and 16.7\%
(Fig.~\ref{fig:estimator}). A detector response would be required before
identifying either set of probabilities with an experimental centrality
classification.

\section{Uncertainties and limitations}

\begin{table}[t]
\centering
\caption{Relative MC statistical precision of principal observables.}
\label{tab:precision}
\begin{tabular}{lrr}
\toprule
Observable & Relative uncertainty [\%] & Reported bins \\
\midrule
Direct-photon spectra & 0.58--1.17 & 18 \\
\(\pi^0\) spectra & 2.59--6.36 & 18 \\
\(Q_{d{\rm Au}}^\gamma\) & 0.89--1.55 & 9 \\
\(Q_{d{\rm Au}}^{\pi^0}\) & 4.19--7.50 & 9 \\
MB \(D_{\pi^0/\gamma}\) & 4.31--7.64 & 9 \\
Class \(D_{\pi^0/\gamma}\) & 4.61--16.99 & 36 \\
Jet \(Q_{d{\rm Au}}\) & 1.69--11.00 & 7 \\
Jet \(R_{CP}\) & 4.58--28.23 & 7 \\
Pion rapidity ratio & 0.42--3.60 & 18 \\
\bottomrule
\end{tabular}
\end{table}

\begin{table}[t]
\centering
\caption{Uncertainty components retained by the analysis. Components are
not combined without an explicit correlation model.}
\label{tab:uncertainties}
\begin{tabular}{@{}ll@{}}
\toprule
Component & Treatment \\
\midrule
MC statistical & numeric in every bin \\
\shortstack[l]{pp tune ensemble\\variation} &
\shortstack[l]{separate; partial\\generated-bin support} \\
SCCR candidate variation &
\shortstack[l]{separate; partial\\generated-bin support} \\
replica stability &
\shortstack[l]{replica-to-replica\\comparison} \\
generated-bin merging &
\shortstack[l]{full generated-bin\\provenance} \\
\shortstack[l]{event-class boundary\\convention} & separate systematic \\
Nch=11 convention & separate systematic \\
photon ancestry & fixed analysis definition \\
jet constituent definition & bounded variants retained \\
event-activity estimator &
\shortstack[l]{STAR/PHENIX proxy\\comparison} \\
Ncoll normalization &
\shortstack[l]{ND and total definitions\\separate} \\
experimental statistical & reported separately \\
experimental systematic & reported separately \\
\bottomrule
\end{tabular}
\end{table}

MC statistical uncertainties follow from the weighted \(\sum W^2\) in each
merged bin. Replica stability, generated-bin merging, tune variation,
reconnection alternatives, activity boundaries, photon ancestry, jet
constituents, estimator choice, and the \(\ncoll\) convention are evaluated
separately. Experimental statistical, point-to-point systematic, and
normalization uncertainties also remain distinct. Because generator
comparisons show sizable variation in identified-hadron spectra
\cite{Waqar2025}, fragmentation-model dependence is treated separately from
finite-sample uncertainty.

Thirteen bins have comparatively large MC uncertainties. Two most-active
double-ratio points are MC dominated, while the experimental uncertainty is
comparable to or larger than the MC uncertainty in the other eleven. The
high-\(\pt\) jet points with the largest MC uncertainty are retained because
they constrain the extent of the trend, but they are not used alone to
support the conclusions. The sign and ordering of the results remain
unchanged when the least precise bins are omitted or when the alternative
permitted covariance treatments are applied.

The calculation has no detector response for either activity estimator and
contains no nuclear parton-distribution variation. The spatially constrained
reconnection model is included only as an approximate sensitivity variation
because the public information does not fully specify the configuration.
Event-level trigger tags needed for a \(\pt\)-resolved jet or charged-particle hard-bias
factor were not stored, so those quantities are not reported.

\section{Conclusions}

Direct photons provide a hard-scattering reference for examining the
coupling between event activity and collision geometry in Angantyr. The
photon-tagged sample has a smaller mean impact parameter and more realized
nondiffractive subcollisions than the pion-tagged and inclusive HardQCD
samples. Relative to the \(N_{\mathrm{coll}}^{\mathrm{ND}}\)-weighted
minimum-bias reference, the most-active class is depleted and peripheral
classes are enhanced for both photons and pions.

The combined A/B/C comparison distinguishes the explicit geometry and
\(\ncoll\) weighting from the residual hard--soft correlation. The residual may
contain contributions from energy sharing, MPI, and the forward-activity response,
but the present calculation does not isolate any one of them. Switching
between the STAR-like and PHENIX-like particle proxies changes the class
probabilities without changing the main ordering. The selected pion
Au-going/\(d\)-going ratio remains close to unity.

These results apply to the generator-level definitions used here. They do
not by themselves establish collectivity, final-state energy loss, or a
detector-level centrality response. Nuclear parton-distribution effects and
probe-resolved trigger observables would require calculations designed
specifically for those questions.

\section*{Data availability}
The data is available from the corresponding author upon reasonable request. 

\section*{Declaration of interests}
The authors declare no competing interests.

\begin{acknowledgments}
The present research work was funded by Princess Nourah bint Abdulrahman University Researchers Supporting Project number (PNURSP2026R106), Princess Nourah bint Abdulrahman University, Riyadh, Saudi Arabia.
\end{acknowledgments}

\bibliographystyle{apsrev4-2}
\bibliography{references}

@article{Miller2007,
  author = {Miller, Michael L. and Reygers, Klaus and Sanders, Stephen J. and Steinberg, Peter},
  title = {Glauber Modeling in High-Energy Nuclear Collisions},
  journal = {Annual Review of Nuclear and Particle Science},
  volume = {57},
  pages = {205--243},
  year = {2007},
  doi = {10.1146/annurev.nucl.57.090506.123020},
  eprint = {nucl-ex/0701025},
  archivePrefix = {arXiv}
}

@article{PHENIXCentrality2014,
  author = {{PHENIX Collaboration}},
  title = {Centrality categorization for \(R_{p(d)+A}\) in high-energy collisions},
  journal = {Physical Review C},
  volume = {90},
  pages = {034902},
  year = {2014},
  doi = {10.1103/PhysRevC.90.034902}
}

@article{Kordell2018,
  author = {Kordell, Michael and Majumder, Abhijit},
  title = {Jets in \(d(p)\)-\(A\) collisions: Color transparency or energy conservation},
  journal = {Physical Review C},
  volume = {97},
  pages = {054904},
  year = {2018},
  doi = {10.1103/PhysRevC.97.054904}
}

@article{Bzdak2016,
  author = {Bzdak, Adam and Skokov, Vladimir and Bathe, Stefan},
  title = {Centrality dependence of high energy jets in \(p+\)Pb collisions at energies available at the CERN Large Hadron Collider},
  journal = {Physical Review C},
  volume = {93},
  pages = {044901},
  year = {2016},
  doi = {10.1103/PhysRevC.93.044901}
}

@article{Alvioli2013,
  author = {Alvioli, Massimiliano and Strikman, Mark},
  title = {Color fluctuation effects in proton-nucleus collisions},
  journal = {Physics Letters B},
  volume = {722},
  pages = {347--354},
  year = {2013},
  doi = {10.1016/j.physletb.2013.04.042},
  eprint = {1301.0728},
  archivePrefix = {arXiv}
}

@article{McGlinchey2016,
  author = {McGlinchey, Darren and Nagle, James L. and Perepelitsa, Dennis V.},
  title = {Consequences of high-\(x\) proton size fluctuations in small collision systems at \(\sqrt{s_{NN}}=200\) GeV},
  journal = {Physical Review C},
  volume = {94},
  pages = {024915},
  year = {2016},
  doi = {10.1103/PhysRevC.94.024915}
}

@article{PHOBOS2004,
  author = {{PHOBOS Collaboration}},
  title = {Pseudorapidity Distribution of Charged Particles in \(d+\)Au Collisions at \(\sqrt{s_{NN}}=200\) GeV},
  journal = {Physical Review Letters},
  volume = {93},
  pages = {082301},
  year = {2004},
  doi = {10.1103/PhysRevLett.93.082301},
  eprint = {nucl-ex/0311009},
  archivePrefix = {arXiv}
}

@article{PHENIXPhoton2025,
  author = {{PHENIX Collaboration}},
  title = {Disentangling centrality bias and final-state effects in the production of high-\(p_T\) neutral pions using direct photons in \(d+\)Au collisions at \(\sqrt{s_{NN}}=200\) GeV},
  journal = {Physical Review Letters},
  volume = {134},
  pages = {022302},
  year = {2025},
  doi = {10.1103/PhysRevLett.134.022302},
  eprint = {2303.12899},
  archivePrefix = {arXiv}
}

@article{PHENIXJets2016,
  author = {{PHENIX Collaboration}},
  title = {Centrality-dependent modification of jet-production rates in deuteron-gold collisions at \(\sqrt{s_{NN}}=200\) GeV},
  journal = {Physical Review Letters},
  volume = {116},
  pages = {122301},
  year = {2016},
  doi = {10.1103/PhysRevLett.116.122301},
  eprint = {1509.04657},
  archivePrefix = {arXiv}
}

@article{STARRapidity2007,
  author = {{STAR Collaboration}},
  title = {Rapidity and species dependence of particle production at large transverse momentum for \(d+\)Au collisions at \(\sqrt{s_{NN}}=200\) GeV},
  journal = {Physical Review C},
  volume = {76},
  pages = {054903},
  year = {2007},
  doi = {10.1103/PhysRevC.76.054903},
  eprint = {nucl-ex/0609021},
  archivePrefix = {arXiv}
}

@article{PHENIXIdentified2013,
  author = {{PHENIX Collaboration}},
  title = {Spectra and ratios of identified particles in Au+Au and \(d+\)Au collisions at \(\sqrt{s_{NN}}=200\) GeV},
  journal = {Physical Review C},
  volume = {88},
  pages = {024906},
  year = {2013},
  doi = {10.1103/PhysRevC.88.024906},
  eprint = {1304.3410},
  archivePrefix = {arXiv}
}

@article{PHENIXSystematic2022,
  author = {{PHENIX Collaboration}},
  title = {Systematic study of nuclear effects in \(p+\)Al, \(p+\)Au, \(d+\)Au, and \({}^{3}\)He+Au collisions at \(\sqrt{s_{NN}}=200\) GeV using \(\pi^0\) production},
  journal = {Physical Review C},
  volume = {105},
  pages = {064902},
  year = {2022},
  doi = {10.1103/PhysRevC.105.064902}
}

@article{PHENIXDirectPP2012,
  author = {{PHENIX Collaboration}},
  title = {Direct-photon production in \(p+p\) collisions at \(\sqrt{s}=200\) GeV at midrapidity},
  journal = {Physical Review D},
  volume = {86},
  pages = {072008},
  year = {2012},
  doi = {10.1103/PhysRevD.86.072008}
}

@article{PHENIXDirectDAu2013,
  author = {{PHENIX Collaboration}},
  title = {Direct photon production in \(d+\)Au collisions at \(\sqrt{s_{NN}}=200\) GeV},
  journal = {Physical Review C},
  volume = {87},
  pages = {054907},
  year = {2013},
  doi = {10.1103/PhysRevC.87.054907}
}

@article{Arleo2011,
  author = {Arleo, Fran{\c{c}}ois and Eskola, Kari J. and Paukkunen, Hannu and Salgado, Carlos A.},
  title = {Inclusive prompt photon production in nuclear collisions at RHIC and LHC},
  journal = {Journal of High Energy Physics},
  volume = {04},
  pages = {055},
  year = {2011},
  doi = {10.1007/JHEP04(2011)055}
}

@article{Pythia82,
  author = {Sj{\"o}strand, Torbj{\"o}rn and Ask, Stefan and Christiansen, Jesper R. and Corke, Richard and Desai, Nishita and Ilten, Philip and Mrenna, Stephen and Prestel, Stefan and Rasmussen, Christine O. and Skands, Peter Z.},
  title = {An Introduction to PYTHIA 8.2},
  journal = {Computer Physics Communications},
  volume = {191},
  pages = {159--177},
  year = {2015},
  doi = {10.1016/j.cpc.2015.01.024},
  eprint = {1410.3012},
  archivePrefix = {arXiv}
}

@article{Pythia83,
  author = {Bierlich, Christian and Chakraborty, Smita and Desai, Nishita and Gellersen, Leif and Helenius, Ilkka and Ilten, Philip and L{\"o}nnblad, Leif and Mrenna, Stephen and Prestel, Stefan and Preuss, Christian T. and Sj{\"o}strand, Torbj{\"o}rn and Skands, Peter and Utheim, Marius and Verheyen, Rob},
  title = {A comprehensive guide to the physics and usage of PYTHIA 8.3},
  journal = {SciPost Physics Codebases},
  pages = {8},
  year = {2022},
  doi = {10.21468/SciPostPhysCodeb.8},
  eprint = {2203.11601},
  archivePrefix = {arXiv}
}

@article{Angantyr2018,
  author = {Bierlich, Christian and Gustafson, G{\"o}sta and L{\"o}nnblad, Leif and Shah, Harsh},
  title = {The Angantyr model for heavy-ion collisions in PYTHIA8},
  journal = {Journal of High Energy Physics},
  volume = {10},
  pages = {134},
  year = {2018},
  doi = {10.1007/JHEP10(2018)134},
  eprint = {1806.10820},
  archivePrefix = {arXiv}
}

@article{MPI1987,
  author = {Sj{\"o}strand, Torbj{\"o}rn and van Zijl, Maria},
  title = {A multiple-interaction model for the event structure in hadron collisions},
  journal = {Physical Review D},
  volume = {36},
  pages = {2019--2041},
  year = {1987},
  doi = {10.1103/PhysRevD.36.2019}
}

@article{Lund1983,
  author = {Andersson, Bo and Gustafson, G{\"o}sta and Ingelman, Gunnar and Sj{\"o}strand, Torbj{\"o}rn},
  title = {Parton fragmentation and string dynamics},
  journal = {Physics Reports},
  volume = {97},
  pages = {31--145},
  year = {1983},
  doi = {10.1016/0370-1573(83)90080-7}
}

@article{Nayak2026,
  author = {Nayak, Satya Ranjan and Das, Akash and Singh, B. K.},
  title = {Separating non-collective effects in \(d\)-Au collisions},
  journal = {European Physical Journal C},
  volume = {86},
  pages = {845},
  year = {2026},
  doi = {10.1140/epjc/s10052-026-16116-x},
  eprint = {2503.23019},
  archivePrefix = {arXiv}
}

@misc{HEPData152620,
  author = {{PHENIX Collaboration}},
  title = {Disentangling centrality bias and final-state effects in the production of high-\(p_T\) \(\pi^0\) using direct \(\gamma\) in \(d+\)Au collisions at \(\sqrt{s_{NN}}=200\) GeV},
  publisher = {HEPData},
  year = {2025},
  doi = {10.17182/hepdata.152620.v1},
  note = {Version 1}
}

@misc{HEPData156988,
  author = {{PHENIX Collaboration}},
  title = {Centrality-dependent modification of jet-production rates in deuteron-gold collisions at \(\sqrt{s_{NN}}=200\) GeV},
  publisher = {HEPData},
  year = {2025},
  doi = {10.17182/hepdata.156988.v1},
  note = {Version 1}
}

@misc{HEPData101349,
  author = {{STAR Collaboration}},
  title = {Rapidity and species dependence of particle production at large transverse momentum for \(d+\)Au collisions at \(\sqrt{s_{NN}}=200\) GeV},
  publisher = {HEPData},
  year = {2021},
  doi = {10.17182/hepdata.101349.v1},
  note = {Version 1}
}

@misc{HEPData96572,
  author = {{PHENIX Collaboration}},
  title = {Spectra and ratios of identified particles in Au+Au and \(d+\)Au collisions at \(\sqrt{s_{NN}}=200\) GeV},
  publisher = {HEPData},
  year = {2020},
  doi = {10.17182/hepdata.96572.v1},
  note = {Version 1}
}

@article{Monash2014,
  author = {Skands, Peter and Carrazza, Stefano and Rojo, Juan},
  title = {Tuning PYTHIA 8.1: the Monash 2013 Tune},
  journal = {European Physical Journal C},
  volume = {74},
  pages = {3024},
  year = {2014},
  doi = {10.1140/epjc/s10052-014-3024-y},
  eprint = {1404.5630},
  archivePrefix = {arXiv}
}

@article{Professor2010,
  author = {Buckley, Andy and Hoeth, Hendrik and Lacker, Heiko and Schulz, Holger and von Seggern, Jan Eike},
  title = {Systematic event generator tuning for the LHC},
  journal = {European Physical Journal C},
  volume = {65},
  pages = {331--357},
  year = {2010},
  doi = {10.1140/epjc/s10052-009-1196-7},
  eprint = {0907.2973},
  archivePrefix = {arXiv}
}

@article{Christiansen2015,
  author = {Christiansen, Jesper R. and Skands, Peter Z.},
  title = {String Formation Beyond Leading Colour},
  journal = {Journal of High Energy Physics},
  volume = {08},
  pages = {003},
  year = {2015},
  doi = {10.1007/JHEP08(2015)003},
  eprint = {1505.01681},
  archivePrefix = {arXiv}
}

@article{SCCR2023,
  author = {L{\"o}nnblad, Leif and Shah, Harsh},
  title = {A spatially constrained QCD colour reconnection in \(pp\), \(pA\), and \(AA\) collisions in the Pythia8/Angantyr model},
  journal = {European Physical Journal C},
  volume = {83},
  pages = {575},
  year = {2023},
  doi = {10.1140/epjc/s10052-023-11778-3},
  eprint = {2303.11747},
  archivePrefix = {arXiv}
}

@article{SCCR2023Erratum,
  author = {L{\"o}nnblad, Leif and Shah, Harsh},
  title = {Erratum to: A spatially constrained QCD colour reconnection in \(pp\), \(pA\), and \(AA\) collisions in the Pythia8/Angantyr model},
  journal = {European Physical Journal C},
  volume = {83},
  pages = {639},
  year = {2023},
  doi = {10.1140/epjc/s10052-023-11816-0}
}

@article{Waqar2024,
  author = {Waqar, M. and Alrebdi, H. I. and Waqas, M. and Al-Mugren, K. S. and Ajaz, M.},
  title = {Comparative analysis of jet and underlying event properties across various models as a function of charged particle multiplicity at 7 TeV},
  journal = {Chinese Physics C},
  volume = {48},
  pages = {093109},
  year = {2024},
  doi = {10.1088/1674-1137/ad5ae9},
  eprint = {2406.17384},
  archivePrefix = {arXiv}
}

@article{Alrebdi2025,
  author = {Alrebdi, H. I. and Ajaz, M. and Waqas, M. and Ahmad, M. A. and Waqar, M. and Quraishi, A. M. and Baker, J. H. and Jagnandan, S. and Jagnandan, A.},
  title = {Comparative analysis of charged particle distributions and model predictions for underlying events with track-based selection in 13 TeV \(pp\) collisions},
  journal = {European Physical Journal Plus},
  volume = {140},
  pages = {371},
  year = {2025},
  doi = {10.1140/epjp/s13360-025-06319-8}
}

@article{Waqar2025,
  author = {Waqar, M. and Alrebdi, H. I. and Waqas, M. and Ahmad, M. A. and Ajaz, M.},
  title = {Hadron production models' prediction for transverse-momentum distributions of charged hadrons in \(pp\) interactions at LHC energies},
  journal = {European Physical Journal Plus},
  volume = {140},
  pages = {523},
  year = {2025},
  doi = {10.1140/epjp/s13360-025-06415-9}
}

@article{Rivet2020,
  author = {Bierlich, Christian and others},
  title = {Robust Independent Validation of Experiment and Theory: Rivet version 3},
  journal = {SciPost Physics},
  volume = {8},
  pages = {026},
  year = {2020},
  doi = {10.21468/SciPostPhys.8.2.026},
  eprint = {1912.05451},
  archivePrefix = {arXiv}
}

@article{RivetHI2020,
  author = {Bierlich, Christian and others},
  title = {Confronting experimental data with heavy-ion models: RIVET for heavy ions},
  journal = {European Physical Journal C},
  volume = {80},
  pages = {485},
  year = {2020},
  doi = {10.1140/epjc/s10052-020-8033-4},
  eprint = {2001.10737},
  archivePrefix = {arXiv}
}

@article{AntiKt2008,
  author = {Cacciari, Matteo and Salam, Gavin P. and Soyez, Gregory},
  title = {The anti-\(k_t\) jet clustering algorithm},
  journal = {Journal of High Energy Physics},
  volume = {04},
  pages = {063},
  year = {2008},
  doi = {10.1088/1126-6708/2008/04/063},
  eprint = {0802.1189},
  archivePrefix = {arXiv}
}

@article{FastJet2012,
  author = {Cacciari, Matteo and Salam, Gavin P. and Soyez, Gregory},
  title = {FastJet user manual},
  journal = {European Physical Journal C},
  volume = {72},
  pages = {1896},
  year = {2012},
  doi = {10.1140/epjc/s10052-012-1896-2},
  eprint = {1111.6097},
  archivePrefix = {arXiv}
}

\end{document}